\documentclass[
 reprint,
 amsmath,amssymb,
 aps
]{revtex4-2}

\usepackage{graphicx}
\usepackage{dcolumn}
\usepackage{bm}
\usepackage[normalem]{ulem}
\usepackage{xcolor}
\usepackage{newtxtext}
\usepackage{lmodern,pdftexcmds}
\usepackage{xstring}
\usepackage{mathrsfs}
\usepackage{soul}

\DeclareMathAlphabet{\mathbfsf}{\encodingdefault}{\sfdefault}{bx}{sl}
\renewcommand{\Vec}[1]{\bm{#1}}
\def\Tens#1{\IfSubStr{ABCDEFGHIJKLMNOPQRSTUVWXYZabcdefghijklmnopqrstuvwxyz}{#1}{\mathbfsf{#1}}{\bm{#1}}}

\newcommand{\figrefp}[2]{Fig.\,\ref{#1}\,(#2)}
\newcommand{\Figrefp}[2]{Figure\,\ref{#1}\,(#2)}
\newcommand{\Appendref}[1]{Appendix\,\ref{#1}}

\definecolor{green}{rgb}{0,0.5,0}

\begin{document}

\preprint{APS/123-QED}

\title{Nonequilibrium orientational bistability of a magnetically-driven slender particle}

\author{Zhiyuan \surname{Zhao}$^{1,\dag,\ast}$}%
\author{Longfei \surname{Li}$^{2,\dag}$}%
\author{Tingting \surname{Sun}$^{1}$}%
\author{Mingcheng \surname{Yang}$^{2,3,\ast}$}%
\author{Masao \surname{Doi}$^{4,}$}%

\email{zhaozhiyuan@ucas.ac.cn; mcyang@iphy.ac.cn; doi.masao@a.mbox.nagoya-u.ac.jp}

\affiliation{
$^1$Department of Physics, Zhejiang University of Science and Technology, Hangzhou, 310023, China}
\affiliation{%
$^2$Beijing National Laboratory for Condensed Matter Physics and Laboratory of Soft Matter Physics, Institute of Physics, Chinese Academy of Sciences, Beijing 100190, China}
\affiliation{%
$^3$School of Physical Sciences, University of Chinese Academy of Sciences, Beijing 100049, China}%
\affiliation{%
$^4$Zhejiang Key Laboratory of Soft Matter Biomedical Materials, Wenzhou Institute, University of Chinese Academy of Sciences, Wenzhou, 325001, China}

\date{\today}


\begin{abstract}

Externally driven systems are non-equilibrium but accessible to stationary states. 
By combining theoretical calculations and particle-based simulations, we investigate the orientational configuration of a transversely-magnetized slender particle in a rotating magnetic field. 
Our results reveal nonequilibrium orientational bistability in both regimes that the particle rotates synchronously and asynchronously with the field.
The relative populations of standing and lying orientations depend strongly on the driving frequency in the synchronous regime, whereas they are primarily controlled by the magnetic field strength in the asynchronous regime.
Effective free-energy approximations further capture the principal features of the steady-state orientational distributions in both regimes. 
Our work aims to give a statistical-thermodynamic description of orientational bistability in magnetically-driven slender particles, paving the way for relevant experimental designs and advanced applications.

\end{abstract}

\maketitle


\section{Introduction}


Externally driven colloidal matter provides a controllable platform for examining how dissipation, fluctuations, and time-dependent forcing reshape configurational order outside thermal equilibrium~\cite{seifert2012stochastic,massana2019tunable,geigenfeind2020superadiabatic,ginot2025energy}. 
Rotating magnetic fields are particularly useful in this context because they impose torques without direct mechanical contact and can be varied in amplitude, frequency, and direction with high precision.
Such a capability has enabled field-programmed assembly~\cite{velez2015magnetic,xie2019reconfigurable}, active transport~\cite{zhang2009characterizing,garcia2018magnetically,pal2018maneuverability,lin2018magnetically}, microrobotic manipulation~\cite{yu2019active,yang2021motion,landers2025clinically}, and rheological probing~\cite{chevry2013magnetic,gu2016ferromagnetic,berret2016local} by magnetic colloids. 
In these systems, anisotropy plays a specific role because the particle shape, magnetic moment, and hydrodynamic resistance generally define different axes of response~\cite{tierno2014recent,thelen2023synergistic}. 
Their competition gives rise to synchronized rotation, nonlinear oscillation, precession, and field selected orientational configurations~\cite{dhar2007orientations,tierno2009overdamped,ghosh2012dynamical,mecke2023simultaneous} that have no direct counterpart in equilibrium suspensions.


Despite extensive progress, predicting the orientational steady states of magnetically-driven slender particles remains challenging. 
Most theoretical work~\cite{ghosh2013analytical,morozov2017dynamics,junot2023unveiling,misra2026dynamics} focus on deterministic dynamics and study hydrodynamic mobility, magnetization models, or step out dynamics of the particles. 
Thermal fluctuations and orientational probability distributions are commonly treated as secondary effects. 
This limitation becomes significant for nanoscale particles, where Brownian motion competes with magnetic alignment and gravitational bias. 
Earlier studies of magnetic nanorods~\cite{dhar2007orientations,ghosh2013analytical,junot2023unveiling} and chiral nanomotors~\cite{morozov2014dynamics} demonstrated that rotating fields can switch the particles between distinct orientational modes, including synchronized and desynchronized rotation around different axes. 
Yet a general thermodynamic understanding of how these modes emerge from the nonequilibrium driving remains largely unexplored.


A key issue is whether the field-driven orientation transition, commonly treated as dynamical bifurcations, can be viewed as thermodynamical competition between nonequilibrium configurational basins. 
These two perspectives provide distinct but complementary information.
A bifurcation-based description identifies deterministic attractors and stability thresholds, whereas a probability-based description can reveal coexistence, noise stabilized states, and transition pathways between competing configurations. 
%
%
Moreover, peaks in the orientational probability distribution need not coincide with deterministic attractors.
Thermal fluctuations can weaken or even suppress the bimodal structure without eliminating the underlying deterministic bistability.
Accordingly, the resulting steady state may contain more than one statistically significant orientation, even when the deterministic dynamics appear simple.


Here, we theoretically investigate the orientational dynamics of a transversely-magnetized slender particle in an external rotating magnetic field by combining Smoluchowski equation, particle-based simulations, and theoretical analysis.
We resolve the steady-state probability distribution and identify how field intensity, driving frequency, thermal noise, and gravity jointly control the transition between lying and standing configurations. 
In both the regimes that the particle rotates synchronously and the asynchronously with the field, we analyze the deterministic dynamical structures that interpret the origin of the orientation transition and, more significantly, the emergence of orientational bistability. 
One-dimensional free-energy-like landscapes are further introduced to capture the principal features of the steady-state orientational distributions in both regimes. 
Finally, we explore the thermal effect on the steady-state orientational distributions.


\section{Methods}
\label{sec_method}


We consider a slender magnetic particle immersed in a viscous fluid and subjected to a spatially uniform magnetic field $\Vec{H}$ rotating in the horizontal ($x$–$y$) plane at angular frequency $\Omega$ [see \figrefp{Figure_1}{a}].
The particle orientation is described by the unit vector $\hat{\Vec{n}}$ along its long axis.
The polar angle $\theta\in[0,\pi/2]$ is measured from the vertical ($z$) axis, while $\phi$ denotes the unwrapped azimuthal angle of the projection of $\hat{\Vec{n}}$ onto the $x$-$y$ plane. 
The magnetic dipole moment $\Vec{m}$ is perpendicular to $\hat{\Vec{n}}$.
Because the field is spatially uniform, the magnetic torque depends only on the net magnetic moment rather than its position along the particle.
Therefore, the model also applies to nonmagnetic particles carrying a transverse magnetic end or coating.


Assuming that the particle rotates about a fixed lower end and has a uniform mass distribution, its gravitational potential energy is given by
\begin{equation}
    U_{\mathrm{g}} 
    = 
    m_\mathrm{eff} g \frac{l}{2} \cos \theta,
    \label{eq:gravity_potential}
\end{equation}
where $m_{\mathrm{eff}}$ is the effective mass, $g$ is the gravitational acceleration, and $l$ denotes the particle length.
The magnetic potential energy is 
\begin{equation}
    U_{\mathrm{m}} 
    = 
    - \mu_0 \Vec{m} \cdot \Vec{H},
    \label{eq:magnetic_potential_origin}
\end{equation}
where $\mu_0$ is the vacuum permeability and $\Vec{m} = m \hat{\Vec{m}}$, with $m$ and $\hat{\Vec{m}}$ denoting the magnitude and orientation of the magnetic dipole moment, respectively.
For a slender particle, we assume that rotation about its long axis relaxes much faster than changes in $\hat{\Vec{n}}$.
Minimizing the magnetic energy over this fast rotational degree of freedom yields
\begin{equation}
    U_{\mathrm{m}} 
    = 
    - \mu_0 m H_0 \sqrt{1 - \sin^2 \theta \cos^2 \varphi},
    \label{eq:magnetic_potential} 
\end{equation}
where $H_0 = |\Vec{H}|$, $\varphi = \Omega t-\phi$ denotes the phase lag in the co-rotating frame, and $t$ is time.
The total potential energy of the system is then $U_{\mathrm{tot}} = U_{\mathrm{g}} + U_{\mathrm{m}}$.


In the overdamped limit, the orientational probability density $\psi(\theta,\varphi;\tilde t)$ evolves according to the Smoluchowski equation on the upper hemisphere~\cite{doi2013soft},
\begin{equation}
    \frac{\partial \psi}{\partial \tilde t} 
    = 
    -\frac{1}{\sin \theta} 
    \frac{\partial}{\partial \theta}
    \left( j_\theta \sin \theta \right)
    -
    \frac{1}{\sin \theta} 
    \frac{\partial j_\varphi}{\partial \varphi},
    \label{eq:smoluchowski_scaled}
\end{equation}
where $\tilde t = t/\tau$ is the dimensionless time, $\tau = 2 \zeta_{\mathrm{r,\perp}}/(m_{\mathrm{eff}}gl)$ is the characteristic rotational relaxation time, and $\zeta_{\mathrm{r,\perp}}$ is the rotational friction coefficient for rotation perpendicular to $\hat{\Vec{n}}$.


The probability fluxes along the $\theta$ and $\varphi$ directions are given by
\begin{gather}
    j_\theta
    =
    -\left(
    T^\ast \frac{\partial \psi}{\partial \theta} 
    +
    \psi \frac{\partial \tilde U_\mathrm{tot}}{\partial \theta} 
    \right),
    \label{eq:flux_theta} 
    \\
    j_\varphi
    =
    - \frac{1}{\sin \theta} \left(
    T^\ast \frac{\partial \psi}{\partial \varphi} 
    +
    \psi \frac{\partial \tilde U_\mathrm{tot}}{\partial \varphi} 
    \right) 
    + 
    \tilde \Omega \psi \sin \theta,
    \label{eq:flux_varphi} 
\end{gather}
where $T^\ast = 2 k_\mathrm{B} T/(m_\mathrm{eff} g l)$ is the dimensionless temperature, $k_\mathrm{B}$ is the Boltzmann constant, $T$ is the absolute temperature, and $\tilde \Omega = \tau \Omega$ is the dimensionless driving frequency.
The dimensionless potential is defined by 
\begin{equation}
    \tilde U_\mathrm{tot} = \cos \theta - h \sqrt{1 - \sin^2 \theta \cos^2 \varphi},
    \label{eq:dimensionless_potential}
\end{equation}
with $h = 2 \mu_0 m H_0/(m_\mathrm{eff} g l)$ denoting the dimensionless magnetic field strength.
The last term in Eq.\,\eqref{eq:flux_varphi} accounts for the rotation of the co-rotating frame.
The probability density is normalized with respect to the spherical measure, such that
\begin{equation}
    \int^{2\pi}_0 \int^{\pi/2}_0
    \psi(\theta,\varphi; \tilde t)
    \sin \theta \, \mathrm{d} \theta \, \mathrm{d} \varphi
    =1.
\end{equation}


We solve Eq.\,\eqref{eq:smoluchowski_scaled} using a cell-centered finite-volume method on a uniform angular grid with $\Delta \theta = \pi / (2 N_\theta)$ and $\Delta \varphi = 2 \pi / N_\varphi$, where $N_\theta$ and $N_\varphi$ are the numbers of grid cells in the polar and azimuthal directions.
Thus, the exact solid-angle weight of cell $(i,k)$ is
\begin{equation}
    V_{ik} = \Delta \varphi \,
    ( \cos \theta_{i-1/2} - \cos \theta_{i+1/2} ).
\end{equation}
where $i$ and $k$ index the polar and azimuthal cells, respectively, and $\theta_{i \pm 1/2}$ denote the polar cell boundaries.
By imposing periodicity in $\varphi$ and setting $j_\theta$ to zero at the boundaries of $\theta = 0$ and $\pi/2$, the discrete scheme conserves the total probability, 
\begin{equation}
    \sum_{ik} V_{ik}\psi_{ik} 
    = 1.
    \label{eq:probability_conservation}
\end{equation}


To maintain stability when potential-driven drift dominates rotational diffusion, we evaluate the probability flux using the Scharfetter-Gummel exponential-fitting scheme
\begin{gather}
    j_f = \frac{T^*}{\Delta s} \left[B(-\mathrm{Pe}_f)\psi_L-B(\mathrm{Pe}_f)\psi_R\right],\\
    B(x) = \frac{x}{e^x-1},
    \label{eq:sg}
\end{gather}
where the subscript $f = \theta$ or $\varphi$, $L$ and $R$ denote the adjacent cells with the positive normal pointing from $L$ to $R$, and $\Delta s$ is the cell separation along the corresponding angular direction. 
For polar faces, the local cell Péclet number $\mathrm{Pe}_\theta = -(\tilde U_R - \tilde U_L) / T^\ast$, whereas for azimuthal faces,
\begin{equation}
    \mathrm{Pe}_\varphi
    =
    \frac{
    -(\tilde U_R - \tilde U_L)
    +
    \tilde \Omega \sin^2 \theta_i \, \Delta \varphi}{T^\ast}.
\end{equation}
Notice that $\tilde U_L$ and $\tilde U_R$ are the dimensionless total potentials at the adjacent cell centers.
The scheme approaches centered diffusion for $|\mathrm{Pe}_f| \ll 1$ and upwind transport for $|\mathrm{Pe}_f| \gg 1$.


The spatial discretization yields
\begin{equation}
    \frac{\mathrm{d} \Tens{\psi}}{\mathrm{d} \tilde t}
    =
    \Tens{A} \Tens{\psi},
\end{equation}
where $\Tens{\psi}$ contains the cell-centered probability densities and $\Tens{A}$ is the sparse conservative operator.
For $T^\ast > 0$, the stationary distribution is obtained by solving $\Vec{A} \Vec{\psi}_\mathrm{st} = \Vec{0}$ together with Eq.\,\eqref{eq:probability_conservation}.
Unless otherwise specified, we use $N_\theta = 80$, $N_\varphi=160$, and $T^\ast = 0.1$.


\section{Results}


\subsection{Steady-state orientational distribution}


\begin{figure}[tb]
\includegraphics[width=0.48\textwidth]{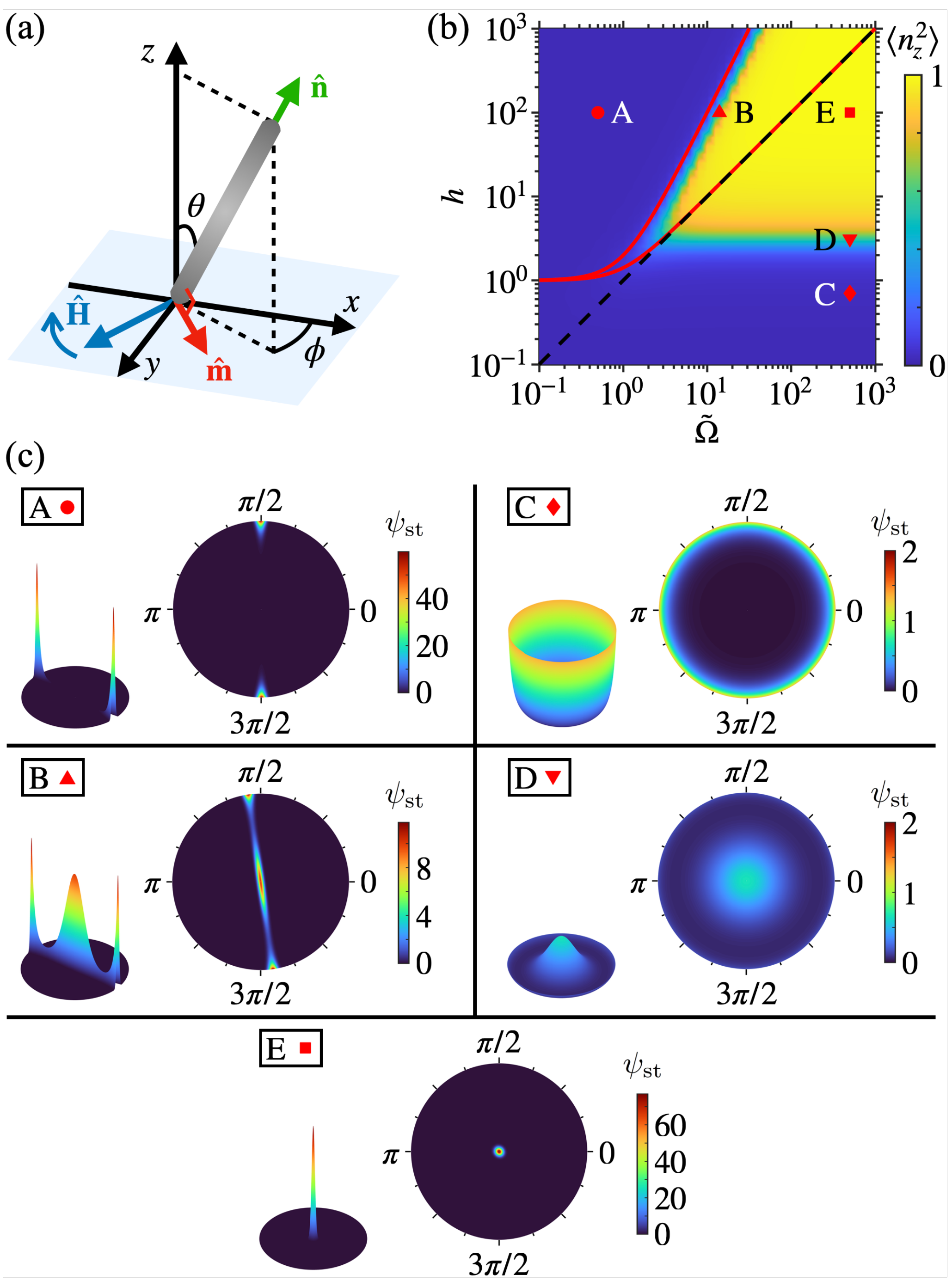}
\caption{
    \label{Figure_1}
    (a) Orientational geometry of a slender magnetic nanoparticle. 
    The unit vectors $\hat{\Vec{H}}$, $\hat{\Vec{m}}$, and $\hat{\Vec{n}}$ denote the directions of the applied rotating magnetic field, magnetic dipole moment, and particle long axis, respectively. 
    The polar angle $\theta$ is measured between $\hat{\Vec{n}}$ and the $z$ axis, and $\phi$ denotes the laboratory-frame azimuthal angle of the projection of $\hat{\Vec{n}}$ onto the $x$-$y$ plane.
    (b) Steady-state map of $\langle n^2_z \rangle$ as a function of the dimensionless field frequency $\tilde \Omega$ and field intensity $h$ at $T^\ast=0.1$. 
    The black dashed line marks $h = \tilde \Omega$, while the domain bounded by the red solid curves indicates $\sqrt{\tilde \Omega^2 + 1} < h < \tilde \Omega^2 + 1$. 
    (c) Top and oblique views of the steady-state orientational probability density per unit solid angle, $\psi_\mathrm{st}(\theta,\varphi)$, for the representative conditions marked in (b). 
    The radial and azimuthal coordinates correspond to $\theta$ and $\varphi$, respectively.
    The parameter sets $(\tilde \Omega,h)$ are
    (A) $(0.5,100)$, 
    (B) $(14,100)$, 
    (C) $(500,0.7)$, 
    (D) $(500,3.12)$, 
    and (E) $(500,100)$. 
    The surface height and color represent $\psi_\mathrm{st}$.
    }
\end{figure}


We evaluate the steady-state orientational order parameter of the particle as
\begin{equation}
    \langle n_z^2 \rangle 
    =
    \int_0^{2\pi} \int_0^{\pi/2}
    \psi_\mathrm{st}(\theta,\varphi) \sin \theta\,\cos^2 \theta \,
    \mathrm{d} \theta \, \mathrm{d} \varphi.
    \label{eq:probability_z}
\end{equation}
The limiting values 0 and 1 correspond to perfectly lying and standing orientations, respectively.
\Figrefp{Figure_1}{b} maps $\langle n_z^2\rangle$ over the $(\tilde\Omega,h)$ parameter plane. 
At low frequencies or weak field strengths, $\langle n_z^2\rangle$ remains close to zero, indicating that the probability is concentrated predominantly at large polar angles $\theta= \pi/2$.
When both $\tilde\Omega$ and $h$ become sufficiently large, $\langle n_z^2\rangle$ approaches unity, indicating prominent localization of the orientational distribution near $\theta = 0$.
However, within an intermediate-frequency window, increasing $h$ first enhances and then suppresses $\langle n_z^2\rangle$, exhibiting a reentrant orientational response.
This result reveals that the steady-state orientation is governed jointly by the driving frequency and field strength.


Since $\langle n^2_z \rangle$ is a global measure of orientational order, states with similar values of $\langle n_z^2\rangle$ may differ substantially in angular localization and multimodal structure.
We further examine the full steady-state orientational probability density $\psi_\mathrm{st}(\theta,\varphi)$.
As shown in \figrefp{Figure_1}{c}, five representative distributions are plotted for the parameter sets A to E marked in \figrefp{Figure_1}{b}.
The radial coordinate represents the polar angle $\theta$, with the center and outer rim corresponding to $\theta = 0$ and $\theta = \pi/2$, respectively.
The azimuthal coordinate denotes $\varphi$, which is defined modulo $2\pi$ and displayed over the interval $[0,2\pi)$.
The surface height and color both denote the magnitude of $\psi_\mathrm{st}(\theta,\varphi)$, with the scale indicated by the individual color bars.
To assist visualization, both top and oblique views are provided. 


Conditions A, B, and E illustrate the effect of increasing $\tilde \Omega$ at fixed $h = 100$. 
At $\tilde \Omega = 0.5$ in condition A, the distribution shows two maxima near $\theta = \pi/2$ and sharply localized in $\varphi$.
They are separated by approximately $\pi$ in azimuth, consistent with the $\pi$-periodic symmetry associated with the particle long-axis orientation.
This strong azimuthal localization indicates a lying configuration of the particle with a well-defined phase lag relative to the rotating field, identifying lying synchronous rotation. 
For condition B with intermediate $\langle n^2_z \rangle$, the distribution develops a maximum near $\theta = 0$ while retaining pronounced maxima near $\theta = \pi/2$. 
Such a multimodal structure suggests the coexistence of standing-like and lying-like orientational populations.
At condition E, $\psi_\mathrm{st}$ collapses into a sharply localized peak near $\theta = 0$, consistent with $\langle n^2_z \rangle \rightarrow 1$ and identifying a robustly standing-dominated orientation. 


Conditions C, D, and E illustrate the effect of increasing $h$ at fixed $\tilde \Omega = 500$.
At condition C with $h = 0.7$, the probability density is concentrated near $\theta = \pi/2$ but nearly uniform over $\varphi$.
The resulting bowl-like distribution indicates a preference for lying orientations with azimuthal delocalization, which is characteristic of lying asynchronous rotation. 
As $h$ increases to $3$ (condition D), the distribution remains nearly azimuthally uniform, while the probability density is concentrated near $\theta = 0$ and weakly near $\theta = \pi/2$.
In comparison with condition B that has a comparable intermediate value of $\langle n^2_z \rangle$, condition D shows much smaller peak-to-background density contrast. 
Upon further increasing $h$ to 100 (condition E), the particle again prefers a predominantly standing orientation.


\subsection{Orientational bistability in the synchronous regime}


\begin{figure}[tb]
\includegraphics[width=0.48\textwidth]{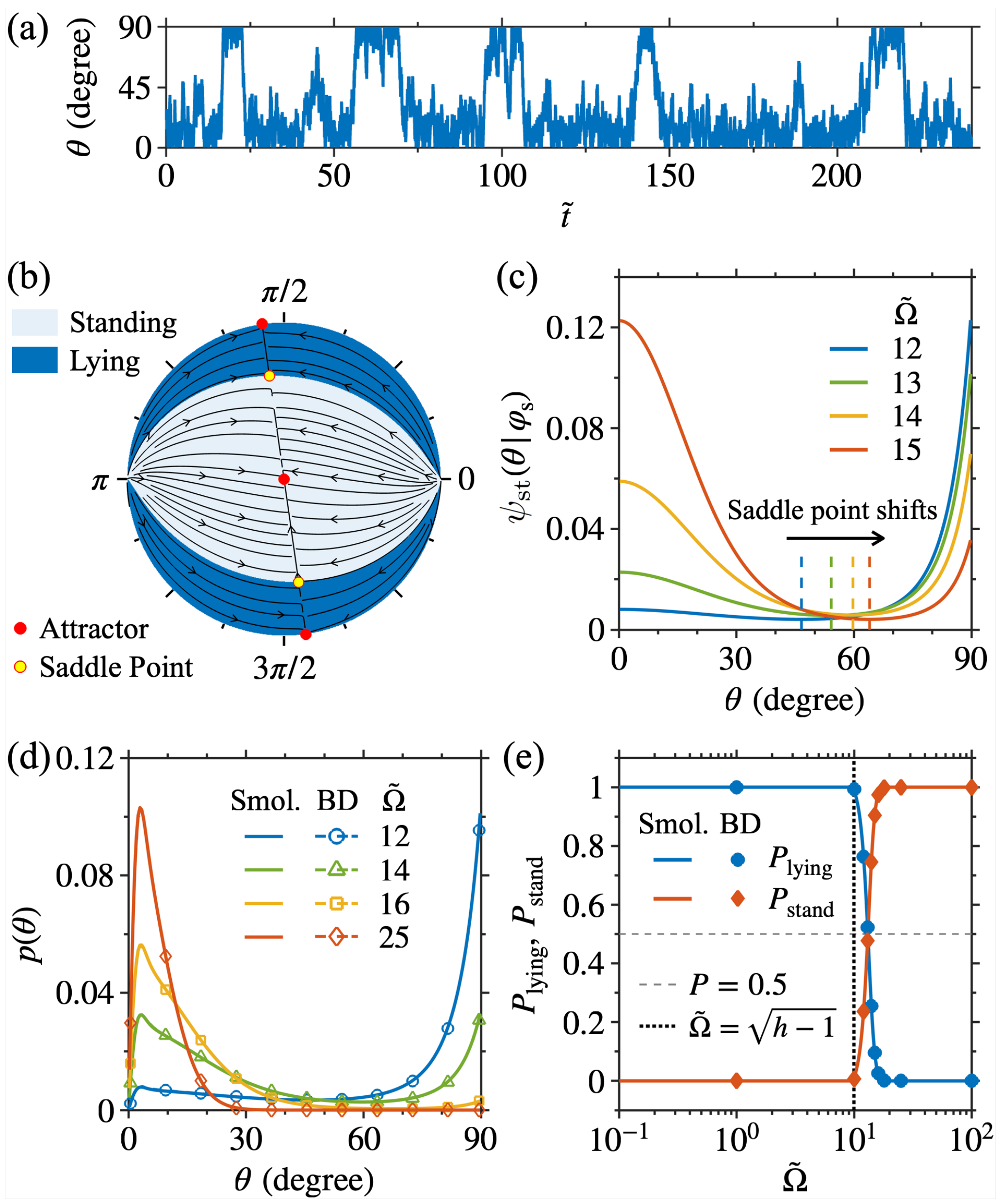}
\caption{
    \label{Figure_2}
    Orientational bistability in the synchronous regime.
    (a) Representative steady-state trajectory of $\theta$ as a function of $\tilde t$, obtained from BD simulations at $\tilde\Omega=14$.
    (b) Deterministic phase portrait in orientational space at $\tilde\Omega=14$. 
    The dark and light shaded regions represent the basins of attraction of the lying and standing orientations, respectively. 
    Red and yellow dots denote attractors and saddle points, respectively.
    (c) Steady-state orientational probability density at the selected azimuthal angle $\varphi_\mathrm{s}$ as a function of $\theta$ for different $\tilde\Omega$, obtained from the Smoluchowski equation. 
    The vertical dashed lines indicate the corresponding saddle-point angles $\theta_\mathrm{s}$ predicted by the deterministic analysis.
    (d) Marginal polar-angle probability density $p(\theta)$ for different $\tilde\Omega$.
    (e) Steady-state occupation probabilities of the lying ($P_\mathrm{lying}$) and standing ($P_\mathrm{stand}$) orientations as functions of $\tilde\Omega$.
    In (d) and (e), the results of the Smoluchowski equation and BD simulations are compared. 
    All panels show $h=100$. 
    The stochastic results are obtained at $T^\ast=0.1$.
    }
\end{figure}


In order to investigate whether the orientational distribution observed in condition B of \figrefp{Figure_1}{c} exhibits dynamical bistability, we perform Brownian dynamics (BD) simulations to obtain the time evolution of the particle orientation.
The details of the BD method is referred to \Appendref{app:BD_method}.
\Figrefp{Figure_2}{a} shows the representative time evolution of the polar angle $\theta$ in the nonequilibrium steady state at $h = 100$ and $\tilde\Omega = 14$. 
One can observe that the particle alternates between orientations near $\theta = 0$ and $\theta = \pi/2$, with intermittent transitions induced by thermal fluctuations. 
This switching behavior provides direct dynamical evidence for the coexistence of two preferred orientational states under identical driving conditions.


The origin of such a bistability is revealed by the deterministic dynamics analysis (see \Appendref{app:deterministic_dynamics}) at the same parameter set.
The phase portrait in \figrefp{Figure_2}{b} shows a standing attractor at $\theta = 0$ and azimuthal-symmetric lying attractors at $\theta = \pi/2$. 
These attracting states are separated by saddle points at 
\begin{gather}
    \theta_\mathrm{s}
    =
    \arcsin \sqrt{\frac{ (\tilde \Omega^2 + 1)^2 - h^2}{\tilde \Omega^2(\tilde \Omega^2 + 1)}},
    \label{eq:saddle_point_theta} 
    \\
    \varphi_\mathrm{s}
    =
    \arccos \left( -\sqrt{\frac{ \tilde \Omega^2 + 1 }{ h^2 }} \right),
    \quad
    \frac{\pi}{2} < \varphi_\mathrm{s} < \pi,
    \label{eq:saddle_point_varphi2}
    \\
    \varphi_\mathrm{s}
    =
    2 \pi - \arccos \sqrt{\frac{ \tilde \Omega^2 + 1 }{ h^2 }},
    \quad
    \frac{3 \pi}{2} < \varphi_\mathrm{s} < 2 \pi,
    \label{eq:saddle_point_varphi1}
\end{gather}
Equation\,\eqref{eq:saddle_point_theta} suggests that either decreasing $h$ at fixed $\tilde \Omega$ or increasing $\tilde \Omega$ at fixed $h$ moves the saddle points from $\theta = 0$ toward $\pi /2$. 
As a result, the statistical weight is redistributed toward the standing basin, accounting for the enhancement yet reentrant behavior of $\langle n^2_z \rangle$ seen in \figrefp{Figure_1}{b}. 



Moreover, the separatrices associated to the saddle points divide the orientation space into distinct basins of attraction of lying and standing orientations, denoted by the dark and light shaded regions, respectively. 
Extending the stability analysis across the $(\tilde\Omega,h)$ parameter plane also yields the boundaries superimposed on \figrefp{Figure_1}{b}. 
The black dashed line separates the synchronous ($h>\tilde \Omega$) and asynchronous ($h<\tilde \Omega$) rotation regimes, while the red solid curves, defined by $h = \sqrt{\tilde \Omega^2 + 1}$ and $h = \tilde \Omega^2 + 1$, delineate the region with coexistence of attractors at $\theta = \pi/2$ and $0$.


\Figrefp{Figure_2}{c} shows $\psi_\mathrm{st}(\theta,\varphi_\mathrm{s})$ at fixed $h=100$ for different $\tilde\Omega$. 
The local minima between the standing-like and lying-like probability maxima closely follow $\theta_\mathrm{s}$ predicted by the deterministic analysis, as indicated by the vertical dashed lines. 
When $\tilde\Omega$ increases, the saddle point shifts progressively from the vicinity of $\theta=0$ toward $\theta=\pi/2$, accompanied by a corresponding displacement of the probability density minimum. 
This shifting as well as the probability density variations at the two attractors explain the significant enhancement of $\langle n^2_z \rangle$ observed in \figrefp{Figure_1}{b}.


We further calculate the steady-state marginal polar-angle probability density by 
\begin{equation}
    p(\theta)
    =
    \sin\theta 
    \int_0^{2\pi} \psi_\mathrm{st}(\theta,\varphi) \, \mathrm{d} \varphi.
    \label{eq:marginal_probability_density}
\end{equation}
Unlike the fixed-azimuth profiles in \figrefp{Figure_2}{c}, $p(\theta)$ accounts for the contributions from all azimuthal angles and provides a normalized measure of the probability per unit polar angle. 
As shown in \figrefp{Figure_2}{d} at $h=100$, the marginal distribution evolves from a lying-dominated state at low $\tilde\Omega$ to a standing-dominated state at high $\tilde\Omega$, with the orientational bistability present at intermediate frequencies. 
At $\tilde\Omega = 25$, the lying-like probability maxima is no longer clearly distinguishable, although the interior saddle points and both orientational attractors still exist.
Additionally, the distributions obtained from the Smoluchowski equation accord well with those extracted from independent BD simulations, offering a quantitative validation of the steady-state probability description.


Based on $p(\theta)$, the relative occupations of the lying and standing orientations are quantified by
\begin{gather}
    P_\mathrm{lying}
    =
    \int^{\pi/2}_{\theta_\mathrm{s}} p(\theta) \, \mathrm{d} \theta.
    \label{eq:P_lying}
    \\
    P_\mathrm{stand}
    =
    \int^{\theta_\mathrm{s}}_0 p(\theta) \, \mathrm{d} \theta, 
    \label{eq:P_stand} 
\end{gather}
\Figrefp{Figure_2}{e} shows that the lying orientation dominates at low frequencies.
As $\tilde \Omega$ increases through the bistable regime, $P_\mathrm{stand}$ rises sharply while $P_\mathrm{lying}$ decreases, indicating a rapid redistribution of probability between the two orientational states. 
The two probabilities become equal near $\tilde\Omega = 13.08$.
At higher frequencies, $P_\mathrm{stand}$ approaches unity, and the standing orientation becomes dominant. 
This redistribution further provides a statistical interpretation of the frequency-dependent orientational response seen in \figrefp{Figure_1}{b}.


\subsection{Orientational bistability in the asynchronous regime}


\begin{figure}[tb]
\includegraphics[width=0.48\textwidth]{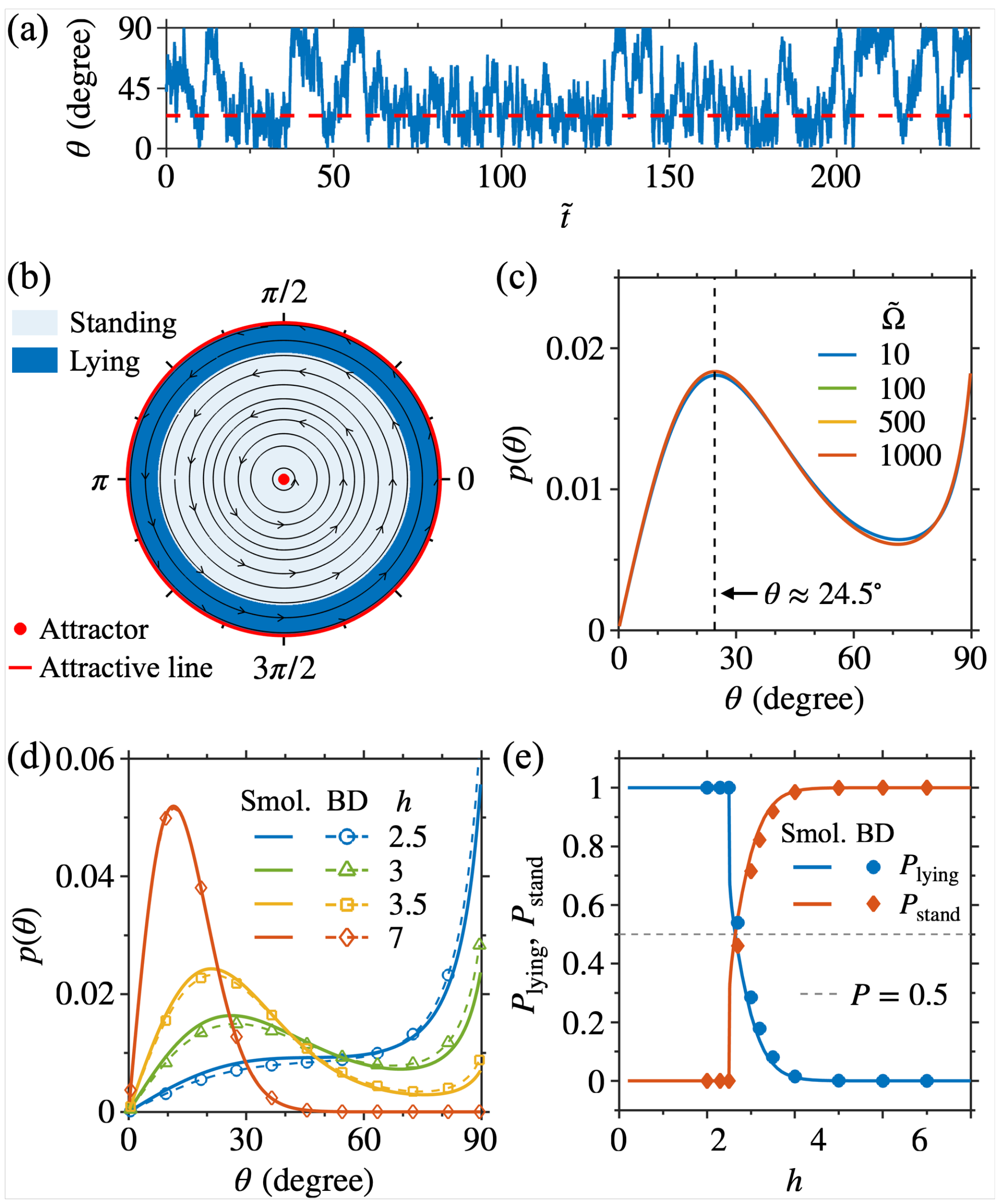}
\caption{
    \label{Figure_3}
    Orientational bistability in the asynchronous regime.
    (a) Representative steady-state trajectory of $\theta$ as a function of $\tilde t$, obtained from BD simulations at $h=3.12$. 
    The red dashed line marks $\theta = 24.5^\circ$.
    (b) Deterministic phase portrait in orientational space at $h=3.12$. 
    The light and dark shaded regions represent the basins of attraction of the standing and lying states, respectively. 
    The red dot denotes the stable standing fixed point, while the red circular line represents the attracting limit cycle associated with the lying state. 
    Marginal polar-angle probability density $p(\theta)$ obtained from the Smoluchowski equation (solid lines) (c) for $h=3.12$ with various $\tilde \Omega$ and (d) for $\tilde \Omega = 500$ with various $h$.
    (e) Steady-state occupation probabilities of the lying-like ($P_\mathrm{lying}$) and standing-like ($P_\mathrm{stand}$) orientations as functions of $h$.
    In (d) and (e), the results of the Smoluchowski equation and BD simulations are compared. 
    Except for (c), all panels correspond to $\tilde\Omega=500$. 
    The stochastic results are obtained at $T^\ast=0.1$.
}
\end{figure}


Orientational bistability is also seen in the asynchronous regime. 
The BD trajectory in \figrefp{Figure_3}{a}, obtained at $\tilde\Omega=500$ and $h=3$ [corresponding to condition D in \figrefp{Figure_1}{c}], exhibits intermittent switching between standing-like orientations at small polar angles ($0<\theta<\pi/4$) and lying orientations near $\theta=\pi/2$. 
Unlike the synchronous case, however, the standing-like orientation undergoes substantial angular fluctuations rather than remaining narrowly localized around $\theta=0$.


The deterministic phase portrait in \figrefp{Figure_3}{b}, obtained from stability analysis (see \Appendref{app:deterministic_dynamics}), reveals a different underlying dynamical structure. 
The standing orientation at $\theta=0$ remains a stable fixed point, whereas the lying state at $\theta=\pi/2$ corresponds to an attracting limit cycle associated with continuous azimuthal rotation. 
Within the two basins of attraction, the trajectories are spiral but appear nearly closed, reflecting the slow polar relaxation relative to the azimuthal motion.
In addition, such basins are separated by an unstable limit cycle that forms a closed separatrix. 
Thermal fluctuations allow the particle to cross this boundary and switch between the two orientational states. 
Such a dynamical structure provides a dynamical basis for interpreting the steady-state distribution observed in condition D of \figrefp{Figure_1}{c}.


\Figrefp{Figure_3}{c} shows that at fixed $h=3.12$, the $p(\theta)$ distributions for various $\tilde \Omega$ nearly overlap, retaining a broad standing-like maximum near $\theta=24.5^\circ$ [also marked by the red dashed line in \figrefp{Figure_3}{a}] and a second maximum near $\theta=\pi/2$.
Our result indicates that the orientational distribution is weakly dependent on the driving frequency in the asynchronous regime. 
It contrasts with the observation reported by Dhar \textit{et al.}~\cite{dhar2007orientations}, where increasing frequency reorients a magnetic rod from standing to lying orientations. 
We attribute this difference to the particle aspect ratio.
For sufficiently slender particles, the rotational resistance about the long axis becomes much smaller than that about transverse axes, leading to the approximated dynamics and then frequency responses shown in the current work. 


The dependence of $p(\theta)$ on $h$ is illustrated in \figrefp{Figure_3}{d} at fixed $\tilde\Omega=500$, where the distributions obtained from the Smoluchowski equation agree quantitatively with those from independent BD simulations.
At $h=2.5$, the probability is concentrated near the lying orientation. 
Increasing $h$ to $3$ produces a broad standing-like maximum at a finite $\theta$ while retaining a probability maximum near $\theta=\pi/2$. 
The standing-like population becomes increasingly pronounced as $h$ increases and dominates at $h=7$. 
The corresponding occupation probabilities in \figrefp{Figure_3}{e} also show a rapid crossover from lying-dominated to standing-dominated orientations as $h$ increases.
Notice that saddle points no more exist in the asynchronous regime, so that $P_\mathrm{lying}$ and $P_\mathrm{stand}$ are estimated based on $\theta$ corresponding to the local minima of $p(\theta)$ instead of $\theta_\mathrm{s}$.
The above results demonstrate orientational bistability in the asynchronous regime and show that the relative populations of the two states depend strongly on the field strength.


\subsection{Effective free-energy description}
\label{sec:effective_potential}


The steady-state orientational distributions in the synchronous and asynchronous regimes can be approximated by effective free-energy landscapes under different dynamical assumptions. 
In the synchronous regime, the phase lag remains localized around its locked value.
This feature motivates a one-dimensional, equilibrium-like approximation for the polar-angle distribution, although the system remains intrinsically driven.


We consider the deterministic torque balance acting on a synchronously rotating particle
\begin{equation}
    \Vec{T}_\mathrm{m}
    +
    \Vec{T}_\mathrm{g}
    +
    \Vec{T}_\mathrm{v}
    =
    \Vec{0}.
    \label{eq:torque_balance} 
\end{equation}
The magnetic torque is $\Vec{T}_\mathrm{m} = \mu_0 \Vec{m} \times \Vec{H}$, while the gravitational torque in the spherical basis is given by $\Vec{T}_\mathrm{g} = (m_\mathrm{eff} g l/2) \sin \theta \, \Vec{e}_\varphi$.
For rigid-body rotation at the imposed angular frequency $\Omega$, the angular velocity can be decomposed as
\begin{equation}
    \Omega \Vec{e}_z
    =
    \Omega (\cos \theta\,\hat{\Vec{n}} 
    -
    \sin \theta \,\Vec{e}_\theta ).
\end{equation}
The corresponding viscous torque is
\begin{equation}
    \Vec{T}_\mathrm{v}
    =
    -\zeta_\mathrm{r,\parallel} \Omega \cos \theta\,\hat{\Vec{n}}
    +
    \zeta_\mathrm{r,\perp} \Omega \sin \theta\,\Vec{e}_\theta,
    \label{eq:viscous_torque} 
\end{equation}
where $\zeta_\mathrm{r,\parallel}$ is the rotational friction coefficient for rotation about the longitudinal axis.


The orthogonality of the gravitational and viscous torque components gives
\begin{equation}
    T^2_\mathrm{m} (\theta, \Omega) 
    =
    \Omega^2 \zeta^2_\mathrm{eff} (\theta)
    +
    \left( m_\mathrm{eff} g \frac{l}{2} \right)^2 \sin^2 \theta,
    \label{eq:torque_balance_square} 
\end{equation}
where $\zeta^2_\mathrm{eff} (\theta) = \zeta^2_\mathrm{r,\parallel} \cos^2 \theta + \zeta^2_\mathrm{r,\perp} \sin^2 \theta$. 
Applying $|\Vec{m} \times \Vec{H}|^2 = m^2 H^2_0 - (\Vec{m} \cdot \Vec{H})^2$ and selecting the magnetic alignment branch with $\Vec{m} \cdot \Vec{H} \ge 0$, we obtain the reduced magnetic energy
\begin{equation}
    U^\ast_\mathrm{m} (\theta, \Omega)
    =
    -\sqrt{\mu^2_0 m^2 H^2_0
    -
    T^2_\mathrm{m} (\theta, \Omega)}.
    \label{eq:reduced_magnetic_energy} 
\end{equation}
This expression is restricted to orientations for which the radicand is non-negative and a consistent phase-locked solution exists. 


The reduced magnetic energy, together with gravity and the spherical measure, motivates an effective landscape
\begin{equation}
    \mathcal{F}_\mathrm{syn} (\theta,\Omega)
    =
    U^\ast_\mathrm{m} (\theta, \Omega)
    +
    m_\mathrm{eff} g \frac{l}{2} \cos \theta
    -
    k_\mathrm{B} T \ln \sin \theta,
    \label{eq:effective_landscape} 
\end{equation}
where the logarithmic term originates from the spherical measure.
By introducing the dimensionless variables and applying Eqs.\,\eqref{eq:torque_balance_square} and \eqref{eq:reduced_magnetic_energy}, Eq.\,\eqref{eq:effective_landscape} can be rewritten as
\begin{align}
    \frac{\mathcal{F}_\mathrm{syn} (\theta,\tilde \Omega)}{m_\mathrm{eff} g l/2}
    =
    - &
    \sqrt{h^2
    -
    \tilde \Omega^2 (\gamma^2 \cos^2 \theta + \sin^2 \theta)
    -
    \sin^2 \theta} \notag \\
    & +
    \cos \theta
    -
    T^\ast \ln \sin \theta,
    \label{eq:effective_landscape_scaled} 
\end{align}
where $\gamma \equiv \zeta_\mathrm{r,\parallel} / \zeta_\mathrm{r,\perp}$.
For the slender-particle approximation considered here, we take $\gamma=0$.


The competition between gravity and the orientation-dependent viscous torque provides a simple explanation for the emergence of standing-like orientations as the driving frequency increases. 
In the low-frequency limit, the viscous contribution is negligible.
For $h \gg 1$, the effective landscape reduces to
\begin{equation}
    \frac{\mathcal{F}_\mathrm{syn} (\theta,\tilde \Omega \rightarrow 0)}
    {m_\mathrm{eff} g l / 2}
    \simeq
    -
    h 
    + 
    \frac{\sin^2 \theta}{2h}
    +
    \cos \theta
    -
    T^\ast \ln \sin \theta.
\end{equation}
The gravitational term and the measure contribution both favor large polar angles, whereas the magnetic contribution weakly opposes this tendency. 
Therefore, the lying orientation dominates when $h$ is sufficiently large.


At finite $\tilde \Omega$, the viscous torque required to sustain synchronous rotation increases with $\sin \theta$ (in the slender-particle limit). 
This raises the effective landscape at large $\theta$ relative to the standing configuration. 
As $\tilde \Omega$ increases, a standing-like minimum can emerge while the lying minimum persists, producing two preferred orientational regions separated by a local maximum of the effective landscape. 
The resulting competition interprets the multimodal distribution observed in the synchronous bistable regime.


Near the standing configuration with $\theta \ll 1$ or, equivalently, $\theta \ll 57.3^\circ$, expanding Eq.\,\eqref{eq:effective_landscape_scaled} gives
\begin{gather}
    \frac{\mathcal{F}_\mathrm{syn} (\theta,\tilde \Omega)}{m_\mathrm{eff} g l / 2}
    \simeq
    1 - h
    +
    A(\tilde \Omega,h) \theta^2
    -
    T^\ast \ln \theta,
    \\
    A(\tilde \Omega,h)
    =
    \frac{{\tilde \Omega}^2 + 1}
    {2 h}
    -
    \frac{1}{2}
    +
    \frac{T^\ast}{6}.
\end{gather}
For $A (\tilde \Omega) > 0$, the landscape exhibits a local minimum at
\begin{equation}
    \theta_\mathrm{syn,p} (\tilde \Omega,h)
    \simeq
    \sqrt{\frac{T^\ast} { 2 A (\tilde \Omega,h)}}.
    \label{eq:theta_at_peak} 
\end{equation}


\begin{figure}[tb]
\includegraphics[width=0.48\textwidth]{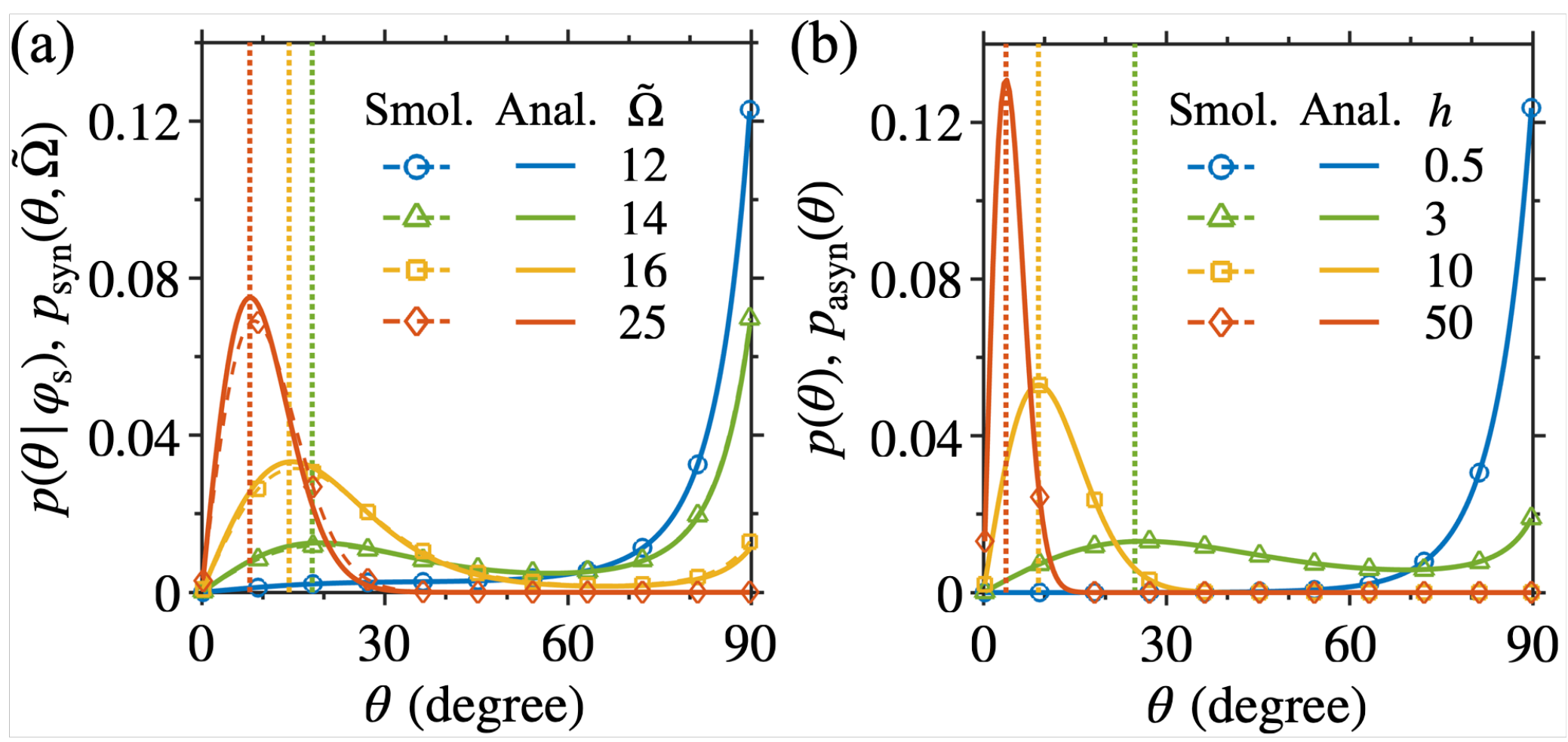}
\caption{
    \label{Figure_4}
    Effective free-energy predictions of the steady-state orientational distributions. 
    (a) Comparison of the conditional probability density $p(\theta|\varphi_\mathrm{s})$ obtained from the Smoluchowski equation with the effective free-energy predictions from Eq.\,\eqref{eq:p_conditional_varphi_at_theta} at $h = 100$ for various $\tilde \Omega$.
    (b) The marginal polar-angle probability density $p(\theta)$ obtained from the Smoluchowski equation and predicted by Eq.\,\eqref{eq:p_marginal_asyn} at $\tilde \Omega = 500$ for various $h$.
    The vertical dotted lines in (a) and (b) indicate the position of the standing-like probability maximum predicted by Eqs.\,\eqref{eq:theta_at_peak} and \eqref{eq:theta_at_peak_asyn}, respectively.
    Both panels show $T^\ast = 0.1$.
}
\end{figure}


Assuming that the reduced polar dynamics admits an approximately Boltzmann-like stationary distribution, we obtain
\begin{equation}
    p_\mathrm{syn}(\theta,\tilde \Omega)
    =
    \frac{\exp[-\mathcal{F}_\mathrm{syn} (\theta,\tilde \Omega) / (k_\mathrm{B} T)]} 
    {\int^{\pi/2}_0 \exp[-\mathcal{F}_\mathrm{syn} (\theta^\prime,\tilde \Omega) / (k_\mathrm{B} T)] \, d\theta^\prime}.
    \label{eq:p_conditional_varphi_at_theta}
\end{equation}
Notice that $p_\mathrm{syn} (\theta,\tilde \Omega)$ describes the polar distribution associated with the phase-locked configuration.
It is different from the polar marginal distribution $p(\theta)$, unless the distribution in $\varphi$ is sharply localized and the probability current along $\theta$ is negligible.
\Figrefp{Figure_4}{a} shows a good agreement between this prediction and $p(\theta|\varphi_\mathrm{s})$ obtained from the Smoluchowski equation at $h=100$ for various $\tilde \Omega$. 
Both the results reproduce the redistribution of probability density from lying to standing-like orientations. 
Equation\,\eqref{eq:theta_at_peak} also well predicts the position of the standing-like maxima, as indicated by the vertical dotted lines.


In the asynchronous regime, $\varphi$ continuously evolves relative to the rotating field.
Especially at sufficiently large $\tilde \Omega$, such a azimuthal motion becomes much faster than the polar relaxation, allowing us to eliminate its angular degree of freedom by taking average over a full rotation. 
Based on Eq.\,\eqref{eq:dimensionless_potential}, the effective one-dimensional landscape can be approximated to 
\begin{equation}
    \frac{\mathcal{F}_\mathrm{asyn}(\theta)}
    {m_\mathrm{eff} g l / 2}
    =
    \frac{1}{2\pi} 
    \int^{2\pi}_0 \tilde U_\mathrm{tot} (\theta, \varphi)\,d\varphi
    -
    T^\ast \ln \sin \theta.
    \label{eq:effective_landscape_asyn} 
\end{equation}
Similarly, near the standing configuration with $\theta \ll 57.3^\circ$, expanding Eq.\,\eqref{eq:effective_landscape_asyn} yields
\begin{gather}
    \frac{\mathcal{F}_\mathrm{asyn} (\theta)}{m_\mathrm{eff} g l / 2}
    \simeq
    1 - h
    +
    C(h) \theta^2
    -
    T^\ast \ln \theta,
    \\
    C(h)
    =
    \frac{h}{4} - \frac{1}{2} + \frac{T^\ast}{6}.
\end{gather}
For positive $C(h)$, this landscape involves a local minimum at
\begin{equation}
    \theta_\mathrm{asyn,p}
    \simeq
    \sqrt{\frac{T^\ast} { 2 C (h)}}.
    \label{eq:theta_at_peak_asyn} 
\end{equation}


The steady-state polar-angle distribution in the asynchronous regime is approximated by a normalized Boltzmann factor as
\begin{equation}
    p_\mathrm{asyn} (\theta)
    =
    \frac{\exp[-\mathcal{F}_\mathrm{asyn} (\theta) / (k_\mathrm{B} T)]} 
    {\int^{\pi/2}_0 \exp[-\mathcal{F}_\mathrm{asyn} (\theta^\prime) / (k_\mathrm{B} T)] \, d\theta^\prime}.
    \label{eq:p_marginal_asyn}
\end{equation}
\Figrefp{Figure_4}{b} manifests $p(\theta)$ obtained from the Smoluchowski equation and predicted by Eq.\,\eqref{eq:p_marginal_asyn} at $\tilde \Omega = 500$.
The agreement extends from the lying-dominated distribution at weak field strengths to the sharply localized standing-like distribution at strong fields, covering the orientational bistable regime. 
The predictions given by Eq.\,\eqref{eq:theta_at_peak_asyn} are also overlapped with the standing-like maxima (see the vertical dotted lines).


\subsection{Thermal effect on the steady-state orientational response}
\label{sec:dependence_temperature}

\begin{figure}[tb]
\includegraphics[width=0.48\textwidth]{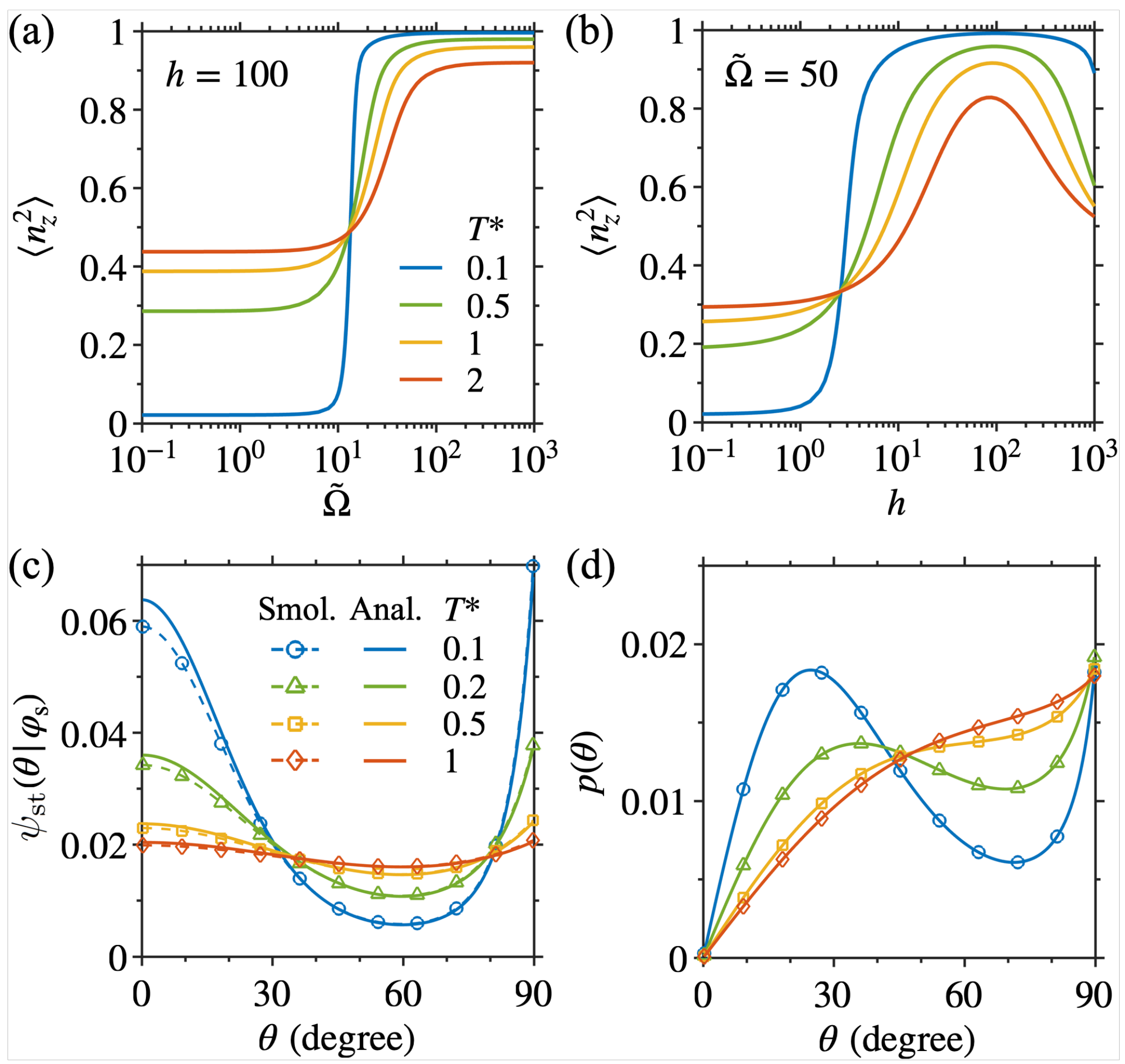}
\caption{
    \label{Figure_5}
    Thermal effect on the steady-state orientational response. 
    Dependence of $\langle n_z^2\rangle$ (a) on $\tilde \Omega$ at fixed $h = 100$ and (b) on $h$ at fixed $\tilde \Omega = 50$ for various $T^\ast$, obtained from the Smoluchowski equation.
    See the legend of (b) in (a). 
    (c) Steady-state orientational probability density at the selected azimuthal angle $\varphi_\mathrm{s}$ as a function of $\theta$ in the synchronous regime at $h = 100$ and $\tilde\Omega = 14$. 
    (d) Marginal polar-angle probability density $p(\theta)$ in the asynchronous regime at $\tilde\Omega = 500$ and $h = 3.12$. 
    In (c) and (d), the result of the Smoluchowski equation is compared with the effective free-energy predictions. 
    See the legend of (d) in (c).
    }
\end{figure}

\Figrefp{Figure_5}{a} and (b) show the orientational order parameter $\langle n_z^2\rangle$ at different $T^\ast$ as functions of $\tilde\Omega$ and $h$, respectively. 
Increasing $T^\ast$ raises $\langle n_z^2\rangle$ in lying-dominated regions while reducing it in standing-dominated regions, indicating the weakening of orientational localization in both configurations. 
At fixed $h=100$, the sharp frequency-dependent reorientation at low temperature becomes progressively broader, with strong standing alignment reached at higher frequencies. 
A similar broadening occurs when $h$ is varied at fixed $\tilde\Omega=50$, where larger field strengths are required to achieve comparable standing alignment as $T^\ast$ increases. 
Particularly, the nonmonotonic dependence of $\langle n_z^2\rangle$ on $h$ persists, although the standing-dominated maximum becomes less pronounced at higher temperatures.
While the deterministic phase boundaries remain unchanged with increasing $T^\ast$, our results suggest that the region of strong standing alignment in \figrefp{Figure_1}{b} shifts toward higher driving frequencies and field strengths, approximately along the $h=\tilde\Omega$ direction.


The influence of thermal fluctuations on the synchronous orientational bistability is further illustrated by the $\psi(\theta|\varphi_\mathrm{s})$ profiles in \figrefp{Figure_5}{c}, where $h = 100$ and $\tilde\Omega = 14$. 
The effective-landscape prediction, obtained by $p_\mathrm{syn}(\theta,\tilde \Omega)/\sin \theta$, accords well with the Smoluchowski result over the examined $T^\ast$. 
Additionally, at low $T^\ast$, the distribution exhibits pronounced maxima near $\theta=0$ and $\theta=\pi/2$, separated by a well-defined minimum. 
Increasing $T^\ast$ reduces the peak-to-valley contrast as probability density redistributes into the intermediate angular region, rendering the bimodal structure less distinct. 
Nevertheless, the angle of the saddle point predicted by the deterministic analysis remains unchanged.
The result indicates that enhanced thermal fluctuation only reduces the contrast between consistent competing orientational basins. 


For the asynchronous regime, \figrefp{Figure_5}{d} exhibits the thermal effect on the marginal polar-angle distribution $p(\theta)$ at $\tilde \Omega = 500$ and $h = 3.12$. 
When $T^\ast$ is low, one can see a standing-like maximum at a finite polar angle and enhanced probability near the lying orientation.
As $T^\ast$ increases, the standing-like maximum broadens and even disappear for $T^\ast \ge 0.5$, while the probability density in the intermediate and large-angle regions increases. 
Such a loss of bimodality indicates that thermal fluctuations increasingly obscure the statistical signature of the coexisting orientational attractors. 
It also supports that stronger magnetic fields are needed to retain the asynchronous orientational bistability at elevated temperatures.


\section{Conclusions}


In this paper, we have investigated the nonequilibrium orientational bistability of a transversely-magnetized slender particle in a rotating magnetic field by combining the theoretical calculations and particle-based simulations.
The steady-state orientational distributions reveal the coexistence of standing-like and lying-like configurations, accompanied by a nonmonotonic orientational response to the driving parameters. 
The underlying bistability exhibits distinct dynamical structures in the synchronous and asynchronous regimes. 
In the synchronous regime, standing and lying orientations correspond to coexisting stable fixed points whose basins of attraction are separated by saddle points, with their relative populations strongly dependent on the driving frequency.
In contrast, the asynchronous bistability arises from the coexistence of a stable standing fixed point and an attracting limit cycle associated with lying rotation, with the relative populations primarily controlled by the magnetic field strength.
In both regimes, thermal fluctuations induce stochastic transitions between the coexisting orientational states.


We further introduce reduced effective-landscape approximations for both synchronous and asynchronous regimes, which capture the principal features of the steady-state orientational distributions.
Moreover, increasing thermal fluctuations broadens the orientational distributions and weakens their bimodal structure. 
These findings provide a clear physical picture of nonequilibrium orientational bistability by linking the underlying dynamical attractors to the stochastic redistribution of orientational populations.
This combined dynamical and statistical description offers a basis for understanding field-controlled reorientation in driven anisotropic particles, with potential implications for responsive colloids and magnetic microrobotic systems.


\begin{acknowledgments}

Z.Z. acknowledge S. Hu and Z. Xiong for valuable discussion. The work was supported by the National Natural Science Foundation of China (No.\,12304247, No.\,T2325027, and No.\,12274448) and startup funds of Wenzhou Institute, University of Chinese Academy of Sciences (No.\,WIUCASQD2020002).
\end{acknowledgments}


\appendix


\section{Rotational Brownian dynamics simulation}
\label{app:BD_method}


We perform rotational Brownian dynamics (BD) simulations by integrating the overdamped rotational equations of motion using a quaternion representation of the particle orientation~\cite{sanchez2009rotational}.
The torques acting on the slender magnetic particle arise from the applied magnetic field, gravity, hydrodynamic resistance, and thermal fluctuations, denoted by $\Vec{T}_\mathrm{m}$, $\Vec{T}_\mathrm{g}$, $\Vec{T}_\mathrm{v}$, and $\Vec{T}_\mathrm{B}$, respectively.
In the absence of rotational inertia, these torques satisfy
\begin{equation}
    \Vec{T}'_\mathrm{m} 
    + \Vec{T}'_\mathrm{g} 
    + \Vec{T}'_\mathrm{v} 
    + \Vec{T}'_\mathrm{B}
    = \Vec{0},
\end{equation}
where the prime indicates vectors in the particle coordinates. 
For an arbitrary vector $\Vec{b}$ in the laboratory coordinates, it can be transferred into the particle coordinates through
\begin{equation}
    \Vec{b}' 
    = 
    \Tens{A} \cdot \Vec{b},
\end{equation}
where $\Tens{A}$ denotes the transformation matrix in terms of quaternions~\cite{evans1977representatation}.


We assume that in the particle coordinates, the permanent dipole moment $\Vec{m}'$ and the particle's long axis $\hat{\Vec{n}}'$ are aligned with the $x'$ and $z'$ axes, respectively.
The magnetic torque is then given by
\begin{equation}
    \Vec{T}'_\mathrm{m}
    =
    \mu_0 \Vec{m}' \times \Vec{H}',
\end{equation}
where the applied magnetic field rotates in the horizontal plane according to
$\Vec{H} = H_0 [ \cos(\Omega t) \Vec{e}_x + \sin(\Omega t) \Vec{e}_y ]$, with unit vectors $\mathbf e_x$ and $\mathbf e_y$ indicating the directions of the laboratory $x$ and $y$ axes, respectively.
Taking the lower end of the rod as the rotational pivot, the gravitational torque is given by
\begin{equation}
    \Vec{T}_\mathrm{g}
    =
    -m_\mathrm{eff} g \frac{l}{2} \hat{\Vec{n}} \times \Vec{e}_z,
\end{equation}
where $\mathbf{e}_z$ is the unit vector along the $z$ axis.
The hydrodynamic resistance torque in the particle coordinates reads
\begin{equation}
    \Vec{T}'_\mathrm{v}
    =
    -\Vec{\zeta}_\mathrm{r} \cdot \Vec{\omega}',
\end{equation}
where $\Vec{\omega}'$ represents the vector of the particle angular velocity and $\Vec{\zeta}_\mathrm{r} = \mathrm{diag}(\zeta_\mathrm{r,\perp}, \zeta_\mathrm{r,\perp}, \zeta_\mathrm{r,\parallel})$ is the rotational resistance tensor.
The Brownian torque is modeled as Gaussian white noise satisfying the fluctuation-dissipation relation
\begin{gather}
    \left \langle
    \Vec{T}_\mathrm{B}'(t)
    \right \rangle
    =
    \Vec{0},
    \\
    \langle \Vec{T}'_\mathrm{B}(0) \Vec{T}'_\mathrm{B}(t)^{\mathsf T} \rangle
    =
    2 k_\mathrm{B} T \Tens{\zeta}_\mathrm{r} \delta(t).
    \label{eq:torque_brownian_covariance}
\end{gather}
where $\delta$ is the Dirac delta function and $\langle \cdot \rangle$ denotes an ensemble average.


Integrating the overdamped angular momentum equation over a time step $\Delta t$ using a first-order Euler method, the infinitesimal rotation vector in the particle coordinates is given by
\begin{equation}
	\Delta \Vec{\Theta}' 
    = 
    \Vec{\zeta}^{-1}_\mathrm{r} \cdot 
    \left( \Vec{T}'_\mathrm{m} + \Tens{A} \cdot \Vec{T}_\mathrm{g} \right) \Delta t 
    + 
    \Vec{W}'(\Delta t),
\end{equation}
where $\Delta \Vec{\Theta}' = [ \Delta \Theta_x', \Delta \Theta_y', \Delta \Theta_z']^\mathsf{T}$ is the rotation increment and $\Vec{W}'(\Delta t)$ is the stochastic rotation increment satisfying 
\begin{gather}
	\langle \Vec{W}'(\Delta t) \rangle
    =
    \Vec{0},
    \\
    \langle \Vec{W}'(\Delta t) \Vec{W}'(\Delta t)^{\mathsf T} \rangle
    =
    2 k_\mathrm{B} T \Tens{\zeta}^{-1}_\mathrm{r} \Delta t.
\end{gather}
The quaternions are updated and re-normalized to unit length after each step~\cite{sanchez2009rotational}.


All the variables are non-dimensionalized according to those shown in the main text.
In the simulations, we set $2\zeta_\mathrm{r,\perp}/(m_\mathrm{eff} g l) = 1000$ and $\zeta_\mathrm{r,\parallel} / \zeta_\mathrm{r,\perp} = 10^{-4}$.
The minimum dimensionless time step is $\Delta\tilde t = 4 \times 10^{-8}$.


\section{Deterministic dynamics analysis}
\label{app:deterministic_dynamics}


In the absence of thermal fluctuations ($T^\ast = 0$), the orientational dynamics of the particle is governed by the potential energy $U_\mathrm{tot}$ and rotational dissipation.
The dissipation function associated with the generalized coordinates $\theta$ and $\phi$ is given by
\begin{equation}
    \Phi
    =
    \frac{\zeta_\mathrm{r,\perp}}{2}
    \left(
    \dot \theta^2 + \dot \phi^2 \sin^2 \theta
    \right),
\label{eq:dissipation_function}
\end{equation}
where $\dot \theta$ and $\dot \phi$ represent the instantaneous rates of change of $\theta$ and $\phi$ in the laboratory coordinates, respectively. 
According to the Onsager variational principle~\cite{doi2013soft}, the equations of motion satisfy
\begin{equation}
    \frac{\partial \Phi}{\partial \dot \theta}
    =
    -\frac{\partial U_\mathrm{tot}}{\partial \theta}
    ,~
    \frac{\partial \Phi}{\partial \dot \phi}
    =
    -\frac{\partial U_\mathrm{tot}}{\partial \phi}.
    \label{eq:Onsager_principle} 
\end{equation}
After transforming the azimuthal dynamics to the co-rotating frame, the deterministic equations of motion read
\begin{gather}
    \tau \dot \theta
    = 
    \sin \theta 
    \left( 1 - 
    \frac{h \cos \theta \cos^2 \varphi}
    {\sqrt{1 - \sin^2 \theta \cos^2 \varphi}} 
    \right),
    \label{eq:evolution_theta} 
    \\
    \tau \dot \varphi
    = 
    \tilde \Omega 
    + 
    \frac{h \cos \varphi \sin \varphi}
    {\sqrt{1 - \sin^2 \theta \cos^2 \varphi}}.
    \label{eq:evolution_varphi} 
\end{gather}
Hereafter, the overdots denote derivatives with respect to $\tilde t$.


The particle can achieve stationary polar angle when $\dot\theta = 0$. 
It gives either the polar branch $\theta = 0$ or the interior branch satisfying
\begin{equation}
    1 - 
    \frac{h \cos \theta \cos^2 \varphi}
    {\sqrt{1 - \sin^2 \theta \cos^2 \varphi}}
    =
    0.
    \label{eq:steady_state_case2} 
\end{equation}
For the polar branch, substituting $\theta = 0$ into Eq.\,\eqref{eq:evolution_varphi} yields
\begin{equation}
    \tau \dot \varphi
    =
    \tilde \Omega 
    +
    \frac{h \sin 2 \varphi}{2}.
    \label{eq:evolution_varphi_case1} 
\end{equation}
When $h \ge 2\tilde\Omega$, phase-locked solutions exist at
\begin{gather}
    \theta_1
    =
    0,
    \label{eq:fixed_point_case1_theta} 
    \\
    \sin 2 \varphi_1
    =
    - \frac{2 \tilde \Omega}{h}.
    \label{eq:fixed_point_case1_varphi}
\end{gather}
For $h < 2\tilde\Omega$, $\dot \varphi$ remains positive for all $\varphi$, indicating continuous drift of $\varphi$. 
Because all values of $\varphi$ at $\theta = 0$ correspond to the same standing orientation, this drift orbit collapses to the pole of the angular space. 
Additionally, within our reduced model with instantaneous spin relaxation, the phase-locked and phase-drifting descriptions on the polar branch are physically indistinguishable in terms of the particle orientation.
Therefore, we focus on the stability of the polar branch rather than on the formal distinction between a fixed point and a degenerate phase drift at $\theta = 0$.


For the interior branch with $\theta \ne 0$, substituting Eq.\,\eqref{eq:steady_state_case2} into Eq.\,\eqref{eq:evolution_varphi} gives
\begin{equation}
    \tau \dot \varphi
    =
    \tilde \Omega 
    +
    \frac{\sin \varphi \cos \varphi}{|\sin \varphi \cos \varphi|} \sqrt{h^2 \cos^2 \varphi - 1}.
    \label{eq:evolution_varphi_case2} 
\end{equation}
Fixed points on this branch appear only in the domain satisfying $\sin \varphi \cos \varphi < 0$ and $\sqrt{\tilde \Omega^2 + 1} < h < \tilde \Omega^2 + 1$.
Their coordinates are given by Eqs.\,\eqref{eq:saddle_point_theta} to \eqref{eq:saddle_point_varphi2} in the main text. 
At the lower and upper limits of this interval, the interior branch approaches the lying boundary and the polar branch, respectively.
Linear stability analysis identifies these interior fixed points as saddle points, whose stable manifolds separate the competing basins of attraction.
Outside the domain where the fixed points exist, $\varphi$ drifts periodically.


The lying configuration corresponds to the artificial boundary $\theta = \pi/2$. 
Along this boundary, Eq.\,\eqref{eq:evolution_varphi} reduces to
\begin{equation}
    \tau \dot \varphi
    = 
    \tilde \Omega 
    + 
    \frac{h \cos \varphi \sin \varphi}
    { |\sin \varphi| },
    \label{eq:evolution_varphi_case3} 
\end{equation}
where $\sin \varphi \neq 0$. 
For $h > \tilde\Omega$, two phase-locked fixed points appear at
\begin{gather}
    \theta_2
    = 
    \frac{\pi}{2},
    \label{eq:fixed_point_case3_theta} 
    \\
    \varphi_2
    =
    \arccos \left( -\frac{\tilde \Omega}{h} \right),
    \quad
    \frac{\pi}{2} < \varphi_2 < \pi.
    \label{eq:fixed_point_case3_varphi2}
    \\
    \varphi_2
    =
    2 \pi - \arccos \frac{\tilde \Omega}{h}, 
    \quad
    \frac{3\pi}{2} < \varphi_2 < 2 \pi,
    \label{eq:fixed_point_case3_varphi1}
\end{gather}
These solutions reach the phase-locking threshold at $h = \tilde\Omega$.
For $h < \tilde\Omega$, no phase-locked lying state exists.
Instead, the particle undergoes continuous azimuthal rotation along the boundary, forming an attracting limit cycle.
When such a lying limit cycle coexists with an attracting standing state, their basins are separated by an unstable limit cycle, as identified from the full two-dimensional phase portrait.


\nocite{*}

\bibliography{library}

\end{document}